\documentclass[conference]{IEEEtran}

\usepackage[utf8]{inputenc}
\usepackage{pdfpages}
\usepackage{todonotes}
\usepackage{lipsum}
\usepackage{hyperref}
\usepackage{hyperref}
\hypersetup{
    colorlinks=true,
    linkcolor=blue,
    filecolor=magenta,      
    urlcolor=cyan,
}
\usepackage{listings}
\usepackage{xcolor}
\usepackage{authblk}
\usepackage{graphicx}
\usepackage{caption}
\usepackage{subcaption}
\usepackage{pdfpages}
 
\definecolor{codegreen}{rgb}{0,0.6,0}
\definecolor{codegray}{rgb}{0.5,0.5,0.5}
\definecolor{codepurple}{rgb}{0.58,0,0.82}
\definecolor{backcolour}{rgb}{0.95,0.95,0.92}

\lstdefinestyle{mystyle}{
    backgroundcolor=\color{white},   
    commentstyle=\color{codegreen},
    keywordstyle=\color{magenta},
    numberstyle=\tiny\color{codegray},
    stringstyle=\color{codepurple},
    basicstyle=\ttfamily\footnotesize,
    breakatwhitespace=false,         
    breaklines=true,                 
    captionpos=b,                    
    keepspaces=true,                 
    numbers=left,                    
    numbersep=5pt,                  
    showspaces=false,                
    showstringspaces=true,
    showtabs=false,                  
    tabsize=2
}
 
\lstdefinestyle{pythonstyle}{
    language=Python,
    basicstyle=\ttm,
    otherkeywords={self},             
    keywordstyle=\ttb\color{deepblue},
    emph={MyClass,__init__},          
    emphstyle=\ttb\color{deepred},    
    stringstyle=\color{deepgreen},
    frame=tb,                         
    showstringspaces=false            %
}

\ifCLASSINFOpdf
\else
\fi

\begin{document}
%
\title{Introducing the Robot Vulnerability Database (RVD)}



\author[1]{Víctor Mayoral Vilches}
\author[1]{Lander Usategui San Juan}
\author[2]{Bernhard Dieber}
\author[1]{\\Unai Ayucar Carbajo}
\author[1]{Endika Gil-Uriarte} 
\affil[1]{Alias Robotics,Vitoria-Gasteiz, Álava, Spain, Email: victor@aliasrobotics.com}
\affil[2]{Institute for Robotics and Mechatronics, JOANNEUM RESEARCH, Klagenfurt am Wörthersee, Austria}

\newcommand\rednote[1]{\textcolor{red}{#1}}






\maketitle

\begin{abstract}
Cybersecurity in robotics is an emerging topic that has gained significant traction. Researchers have demonstrated some of the potentials and effects of cyber attacks on robots lately. This implies safety related adverse consequences causing human harm, death or lead to significant integrity loss clearly overcoming the privacy concerns in classical IT world. \\
In cybersecurity research, the use of vulnerability databases is a very  reliable tool to responsibly disclose vulnerabilities in software products and raise willingness of vendors to address these issues. In this paper we argue, that existing vulnerability databases are of insufficient information density and show some biased content with respect to vulnerabilities in robots. This paper presents the Robot Vulnerability Database (RVD), a directory for responsible disclosure of bugs, weaknesses and vulnerabilities in robots. This article aims to describe the design and process as well as the associated disclosure policy behind RVD. Furthermore the authors present preliminary selected vulnerabilities already contained in RVD and call to the robotics and security communities for contribution to the endeavour of eliminating zero-day vulnerabilities in robotics.

\end{abstract}

\section{Introduction and background}
\label{intro}
A vulnerability is a mistake in software or hardware that can be directly used by an arbitrary malicious actress to gain access to a system or network, operating it into an undesirable manner\cite{Pfleeger:2002:SC:579149}. In robotics, security flaws such as vulnerabilities are of special relevance given the physical connection to the world that these systems imply. As discussed in \cite{kirschgens2018robot}, "\emph{Safety cares about the possible damage a robot may cause in its environment, whilst security aims at ensuring that the environment does not disturb the robot operation.  Safety and security are connected matters. A security-first approach is now considered as a prerequisite to ensure safe operations}".\\
\newline
Robot vulnerabilities are indeed potential attack points in robotic systems that can lead not only to considerable losses of data but also to safety incidents involving humans. Some claim\cite{zheng2011ivda} that unresolved vulnerabilities are the main cause of loss in cyber incidents. The mitigation and patching of vulnerabilities has been an active area of research\cite{ma2001sharing, ALHAZMI2007219, Shin2011Vulnerabilities, Finifter2013BugBounty, McQueen2009Zeroday, Bilge:2012:BWK:2382196.2382284} in computer science and other technological domains. Unfortunately, even with robotics being an interdisciplinary field composed from a set of heterogeneous disciplines (including computer science), not much vulnerability mitigation research has been presented so far. 
\newline
A variety of vulnerability archives and bug-tracking sites exist already. These databases generally provide information that allows security researchers to locate, mitigate or fix flaws in their systems. Arguably, the most popular of such databases is the U.S. National Vulnerability Database (NVD)\cite{booth2013national}, a U.S. government funded repository of vulnerabilities compiled following a series of U.S. guidelines and standards. NVD presents an archive with vulnerabilities, each with their corresponding Common Vulnerabilities and Exposures (CVE)\footnote{Common Vulnerabilities and Exposures (CVE) List CVE® is a dictionary of entries—each containing an identification number, a description, and at least one public reference for publicly known cybersecurity vulnerabilities. CVE contains vulnerabilities and exposures and to date is sponsored by the U.S. Department of Homeland Security (DHS) and by the Cybersecurity Infrastructure Security Agency (CISA) although it is not a database \emph{per se} (see \href{https://cve.mitre.org/about/faqs.html}{official information}). CVE it self does not contain the information in a database manner, but instead, CVE List feeds vulnerability databases (such as the National Vulnerability Database (NVD)) with its entries, and acts as an aggregator of vulnerabilities and exposures reported at NVD.} identifiers. Thus, NVD gets fed by the CVE List and then builds upon the information included in CVE Entries to provide enhanced information for each entry such as fix information, severity scores, and impact ratings.
\newline
There are numerous vulnerability databases, both public and private, however to the best of our knowledge, none of these databases includes more than a few dozens of robot-related vulnerability entries. Moreover, from our research, in most cases, the information provided does not accurately facilitate the reproduction of the flaws or its mitigation since reporting schemes were not originally thought for robotics. In robotics, system integration and the context become critical key factors for the reproduction of any flaw or mitigation. 
\\
\\
When reviewing prior work on vulnerability databases in the context of robotics, this paper identifies the following aspects that deserve further discussion:

\begin{itemize}
    \item \textbf{CVE robot-related results are scarce}: At the time of writing, the current CVE List provides some \emph{humble} results when searching for \emph{robot} (43 CVE entries), \emph{Robot Operating System} (892 CVE entries though most, not robotics related)  or even the misleading \emph{ROS} query (14 CVE entries). A closer look into the results led the research towards realizing that information is scarce, unstructured and in most cases insufficient for vulnerability assessment. Finding robotics-related flaws with the required accuracy is currently challenging. A similar exercise was reproduced in other archives of vulnerabilities with similar results.

    \item \textbf{CVE reports require more details to be used (in robotics)}: Taking  CVE-2019-13566 as an example. Except for the description and a few code pointers, this particular entry provides very little information to help a system integrator or manufacturer determine its relevance. No details regarding the "system under test" have been provided, neither a exploit that confirms its exploitability or a vector that allows to measure its severity according to CVSSv3. Coming from the same group, we find CVE-2019-13465 which at the time of writing is presented as ** RESERVED ** though authors behind it\footnote{Refer to \url{https://bit.ly/35FBcna}} already disclosed that it is related to a \emph{potential iterator cause buffer overflow}. Similarly, CVE-2019-13445 presumably published by the same researchers remains classified. For a test engineer or security researcher aiming to reproduce and assist in patching these flaws, more information is required. The intrinsic system integration of the robotics field demands for additional context such as the version of the robot, or robot component under test (in this case, the Robot Operating System (ROS)\cite{quigley2009ros}). Examples of additional information required may include a well defined, context-specific and appropriate severity scoring mechanism (to prioritize flaws) or a exploit to validate its type and classification.
    
    \item \textbf{Encouraging triage appears of utmost relevance}: Robotics is the art of system integration. Its modular characteristic by nature, both in hardware and software aspects, provides unlimited flexibility to its system designers. This flexibility however comes at the cost of complexity. The qualification of a security flaw commonly known as "triage" seems of special relevance in the domain of robots given its complexity. Establishing a channel that favours an open discussion, where other researchers might contribute is to us beneficial.

    \item \textbf{Assisted  reproduction of flaws}: Working with robots is generally very time consuming. From the authors' experience and involvement in the constructions of robots, it’s an inherent characteristic of the complexity of the field and the trade-off obtained with its modularity. Mitigating a vulnerability or a bug requires one to first reproduce the flaw. This can be extremely time consuming, specially ensuring an appropiate enviroment for its reproduction. The authors consider that it would be beneficial to include on each flaw ticket items that facilitate native Operating System (OS) virtualization via technologies like Linux Containers. By using a technology like Docker \cite{merkel2014docker}, researchers will obtain relevant support in reproducing the flaws leading to faster mitigations.

    \item \textbf{Unfit severity scoring mechanism}: CVE uses the Common vulnerability scoring system (CVSS) \cite{mell2006common} to report on the severity of vulnerabilities. As previously discussed \cite{vilches2018towards}, CVSSv3 has strong limitations when applied to robotics. Simply put, it fails to capture the interaction that robots may have with their environments and humans.

\end{itemize}

Vulnerabilities played a significant role in past attacks affecting other areas \cite{mcmillan2010siemens} and can be judged as the major cause for losses. Specially, the so called \emph{Zero-day} (also known as \emph{0-day}) vulnerabilities, security flaws that are unknown to, or unaddressed by, those who should be interested in mitigating the vulnerability. Conceivably, provided vendors released security patches for vulnerabilities promptly after discoveries, 0-day attacks as well as other attacks and damages using these flaws would be significantly reduced. This demands for manufacturers to be informed about new flaws affecting their systems. However, according to past research \cite{kirschgens2018robot}, most vendors in robotics are currently ignoring security flaws completely. Within the security community, it's commonly accepted that "creating pressure" towards more reasonably-timed fixes results in smaller windows of opportunity for malicious actors to abuse vulnerabilities. Several projects including Project Zero\footnote{\href{https://googleprojectzero.blogspot.com/}{https://googleprojectzero.blogspot.com/}} from Google or the Zero Day Initiative\footnote{\href{https://www.zerodayinitiative.com/}{https://www.zerodayinitiative.com/}} from Trend Micro have adopted this philosophy defining disclosure policies with a maximum time deadline for manufacturers to provide a fix before publicly disclosing the vulnerability. Similar to some of these initiatives, the authors believe that vulnerability disclosure is a two-way street where both vendors and security researchers, must act responsibly.

As described by Zheng et al. \cite{zheng2011ivda}, attempts to resolve this dilemma have resulted in the development of vulnerability disclosure policies. The disclosure of a vulnerability is the revelation of a vulnerability to the public at large.

The authors acknowledge that one of the most---if not the most---important task in security and particularly vulnerability management is minimizing the \emph{time window of vulnerability}. On this regard and in an attempt to provide robot manufacturers and users a valuable source of information, we design and construct a vulnerability database, the Robot Vulnerability Database or RVD for short. Together with RVD, and aiming to reduce 0-days from robotics, we also present an attached disclosure policy thought for robot-related flaws that minimizes the \emph{time window of vulnerability}. This paper aims to describe our approach and discuss our design decisions. The rest of the paper is organized as follows. Section \ref{rvd} introduces RVD and our design choices. Section \ref{results} presents some vulnerability results and  section \ref{bias} argues about our bias in the compilation of such results. Finally, section \ref{conclusion} finalizes with a series of conclusions and future work items.

\section{The Robot Vulnerability Database (RVD)}
\label{rvd}

The Robot Vulnerability Database (RVD) is a database for robot vulnerabilities and bugs that aims to record and categorize flaws that apply to robots and robot components, including software and hardware. The database is freely available at \url{https://github.com/aliasrobotics/RVD} and an open source set of tools to manage the database are also available\footnote{Undocumented at the time of writing} within that same repository.\\
\newline
As first introduced by Ma et al. \cite{ma2001sharing}, this paper discusses the design of the robot vulnerability database by arguing on a set of relevant items.

\subsection{Scope}
The scope of RVD comprises all robotics hardware and software systems, including complete robots but also individual components. 

\subsection{Language and terminology}

Information sharing becomes difficult without a common language. The available vocabulary to discuss computer security concepts is limited which leads to an “overloading” of terms, i.e., a reuse of the same terms with varying scope and level of abstraction. This was observed while reviewing different and existing databases which not only overloaded but also mixed terms such as \emph{weakness}, \emph{bug} and \emph{vulnerability} leading to confusion and misunderstandings.

RVD attaches itself to common language and standards as defined by CVE List including the definitions of weakness, vulnerability and exposure. In addition, we use the term \emph{"flaw"} to refer to all security-related errors. The authors however found somewhat troubling that there was no consensus across security organizations to define when a vulnerability is a vulnerability, and not a bug. This paper argues that this aspect is connected with the lack of resources for reproducing reported vulnerabilities in most databases and thereby accepting the so called "theoretical vulnerabilities" more than usually. Moreover, we question whether this conservative approach of not only "disclosing selectively" but also "disclosing scarcely" can be justified after recent results \cite{ma2001sharing, zheng2011ivda, meunier1999final, christey2013buying, antrobus2016simaticscan, brief2013vulnerability}.\\
\newline
To ensure uniqueness of all robot-related flaws, unique identifiers from CVE List are re-used within RVD and tagged as "cve". Additionally, a unique and iterative identifier "id" is assigned to each new flaw. To categorize flaws, the Common Weakness Enumeration (CWE) is used.

To ensure future growth and adaptation, no further constrains have been applied (e.g. title naming convention). For further clarification on vocabulary, authors refer readers to appendix \ref{appendixA} where terminology is further discussed.

\subsection{Sharing model}
\label{subsec:sharing_model}
To understand the rationale of the data sharing model we refer the readers to the 2nd Workshop of Research with Security Vulnerability Databases \cite{meunier1999final}. In their report, Meunier et al. argue about 4 different models for vulnerability data sharing, namely "Fully Available", "Centralized", "Federated" and the "Balkan/Status Quo". In this paper, It is  briefly described and discussed each one of those models below before introducing our approach:
\begin{itemize}
    \item \textsc{Fully Available}: Characterized by openness. The database is completely open and anyone can access it or add to it. Copies can be made and used freely. This model offers the greatest use of access and eliminates the need of logging activity or authenticating users. Users can download the entire database seamlessly. The biggest disadvantage of this model is that the funding of its maintenance has to be sought.
    \item \textsc{Centralized}: Would entail a database of vulnerabilities
    managed by some central organization that would be in charge of defining the schema, data review and consistency, funding, and policy. Access to the database typically requires some sort of user subscription and authentication. Advantages of this approach are mainly the overall consistency and data quality control. The downside is the scalability and organizational bias introduced by the managing organization.
    \item \textsc{Federated}: as in a loose union of several distributed entities on a common task. Consortium, foundations and similar operate in this manner defining a steering committee. This model distributes responsibilities among potentially qualified parties and ensures funding however risks inequality by favouring partners of the federation with more resources (e.g., big companies).
    \item \textsc{Balkan/Status Quo}: Implies that each "balkan" or participant has their own database of vulnerabilities which she or he is not willing to share with the rest.
\end{itemize}

Out of the research performed, and similar to \cite{meunier1999final}, It has been concluded that everybody is interested in vulnerabilities including software vendors, consumers, security researchers, malicious actresses, foreign governments, terrorists, etc..., whether or not they would be willing to admit it publicly. Objecting to the public distribution of vulnerabilities or failing to acknowledge is effectively a proof of security-immaturity of the players involved. This includes robot or robot component manufacturers embarrased by flaws, pressured by their clients or unable to cope with the security community.\\
\newline
RVD adheres to the \textsc{Fully Available} model for the most part. This project hosts the database in Github which requires no access control for consulting the information, but demands it for contributions or extensions of any kind. This is enforced to a) ensure a standard format of submissions, b) favour the ease of use and c) motivate for-profit entities to give back and contribute, generating credit, credibility and costing them less than maintaining their own database. This project proposes a GPLv3 license for the tools and related-content to ensure enhancements and contributions on top are feed back to the project.\\
\newline
By adopting this model, the falsification and erasure of records controlled by a central entity becomes hard, because valid copies may be saved by anyone exporting the tickets and records. Moreover, the setup proposed, in our best intention, provides great fault-tolerance due to the ease of making non-confidential mirrors and duplicate copies.\\
\newline
In addition, and to empower privacy in advisories to manufacturers or other interested parties, we leave the door open for the integration of RVD with private (non-open) sources of information. We prototyped and deliver a proof of concept using a private source hosted in Gitlab\footnote{Refer to \url{https://bit.ly/2slL3QE}}.

\subsection{Taxonomy}
In their report, Meunier et al. already highlighted that data sharing will remain difficult (expensive) as long as there is no agreement on what is relevant vulnerability data. This easily leads to the need to define a common taxonomy for vulnerabilities and a matching data exchange schema.

Both, the taxonomy and a matching schema are available within the repository. The taxonomy extends prior work related to the classification of bugs in robotics, namely the robust project\footnote{See \url{https://github.com/robust-rosin/robust}}. The schema is available at \url{https://github.com/aliasrobotics/RVD/blob/master/rvd_tools/database/schema.py}. It has been implemented using a simple and easy to extend Python dictionary and enforced using the  \emph{cerberus} library.

\subsection{Access control}

The authors acknowledge that the way in which the information describing vulnerabilities is handled is extremely important. Vulnerability data is very sensitive and therefore should be carefully disclosed. We propose a model for RVD that implements access control for making contributions. By favouring an authenticated disclosure, we hope to favour responsible coherent actions. To simplify and lower the overhead, we rely on Github's native accounting. New tickets are tagged with a "triage" label and RVD maintainers collaborate to triage them out at their earliest availability.

\subsection{User interface}

Similar to Access Control, User Interface relies heavily on Github's native features. By leveraging Github's front-end, RVD gets access to a well reviewed and tested front-end designed for collaboration and participation.

\subsection{Review process}

Each flaw is subject to be reviewed at any point in time. The severity of each flaw is calculated using two scoring mechanisms: the Common Vulnerability Scoring System (CVSS) and the Robot Vulnerability Scoring System (RVSS) \cite{vilches2018towards}. The later is the result of reviewing CVSS for the domain of robotics. RVD implements both, CVSSv3 to ensure compatibility with other databases and RVSS, to provide additional useful information in the robotics context.\\
\newline
Maintainers associate each security flaw to a Github issue. By leveraging Github issues and particularly, the e-mail like interaction, we advocate for assisted flaw review, reproduction and ultimately, assisted triage. Overall, management of the tickets is coordinated by a set of maintainers selected from the contributors of the database. The review of flaws is supported by active use of labels which feed with information to the Continuous Integration and Deployment (CI/CD) system.

\subsection{Maintenance}

As discussed in section \ref{subsec:sharing_model}, funding is required to maintain the database. In order to reduce the financial needs as well as to ensure maintainers are relieved from the more dull tasks, a series of automations are programmed into RVD. Using CI/CD infrastructure, autonomous and semi-autonomous tasks are introduced into RVD and implemented when applicable as Github Actions.\\
\newline
At the time of writing, with different degrees of automation and maturity, the database provides the following features to simplify maintenance:
\begin{itemize}
    \item It produces automatically updated reports (in the form of a README.md) about its status upon new changes. 
    \item It checks and validates conformance of new entries (or existing ones subject to changes) with the schema
    \item It analyzes the database for duplicates using a scalable fuzzy matching library that implements regularized logistic regression augmented with Active Learning. Simply put, it allows to train models that accurately identify duplicates.

\end{itemize}

The authors are actively working on new additions to this list. This should further reduce to overall effort to maintain RVD.

\subsection{Disclosure Policy}

Coherently and to ensure timely responses from manufacturers, together with RVD the authors propose a disclosure policy. Unless specified, authors adhere to a 90-day public disclosure policy for new vulnerabilities after the first communication with the vendor. Full text of the policy is available on RVD's official repository. \\
\newline
This disclosure policy is heavily inspired by Google's Project Zero and is recommended to all maintainers and contributors to RVD. This paper calls on all contributing researchers to adopt disclosure deadlines in some form.

\section{Vulnerability statistics and results}
\label{results}
With the aim of preliminarily discussing our RVD use and feeds, we provide some further data on the entries held at the date of publication of the present work. Table \ref{tab:vuln_vendor} presents a compilation of vendors and the number of vulnerabilities registered within RVD.

\renewcommand{\arraystretch}{1.5} 
\begin{table}[h!]
    \begin{center}
        \begin{tabular}{l c}
        
         \textbf{Vendor} & \textbf{count}  \\ \hline
         ABB & 61  \\ \hline
         Fanuc & 6 \\ \hline
         Robotics & 2 \\ \hline
         Universal Robots & 5 \\ \hline
         DDS vendors (eProsima, ADLINK, RTI) & 2 \\ \hline
         Acutronic Robotics & 5 \\ \hline
         Vecna & 6 \\ \hline
         WowWee & 3 \\ \hline
         UBTech Robotics & 3 \\ \hline
         PAL Robotics & 1 \\ \hline
         SoftBank Robotics & 4 \\ \hline
         Rethink Robotics & 3 \\ \hline
         Asratec & 1 \\ \hline
        \end{tabular}
        \caption{Number of vulnerabilities contained in RVD classified by vendor.}
        \label{tab:vuln_vendor}
    \end{center}
\end{table}

\begin{figure}[h!]
    \includegraphics[width=0.5\textwidth]{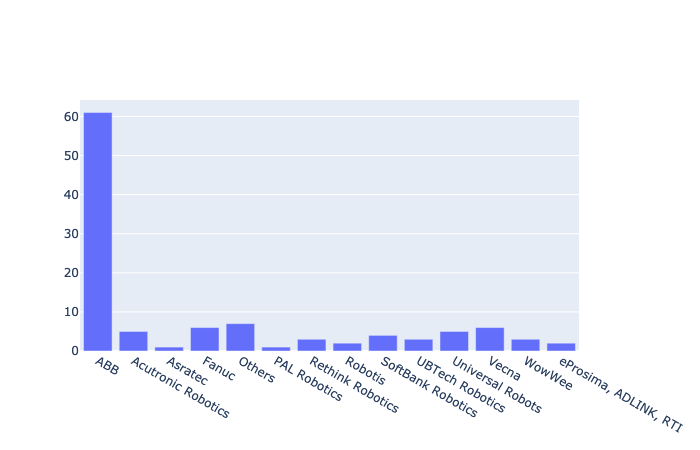}
    \centering
    \caption{Number of vulnerabilities recorded in RVD per manufacturer}
    \label{fig:vendors_count1}
\end{figure}

The total number of vulnerabilities per manufacturer provides some insights. From the raw numbers one can tell that ABB has faced as many vulnerabilities in their robotic systems as the rest of the other manufacturers combined. This aspect is illustrated in Figure \ref{fig:vendors_count1} and could lead one to think ABB's commitment with security far exceeds other vendors. It must be noted however that several of these flaws have not been addressed fully. For example, if one was to consider \href{https://github.com/aliasrobotics/RVD/issues/729}{RVD\#729}, it will be noticeable that the mitigation provided involves stressing certain risks that the user gets exposed to, but no actual mitigating change or update has been made effective. In addition, several of the flaws catalogued for ABB are OT-related and a closer look into them is required.

\begin{figure*}[h!]
    \begin{subfigure}{.48\textwidth}
    \centering
        \includegraphics[width=\textwidth]{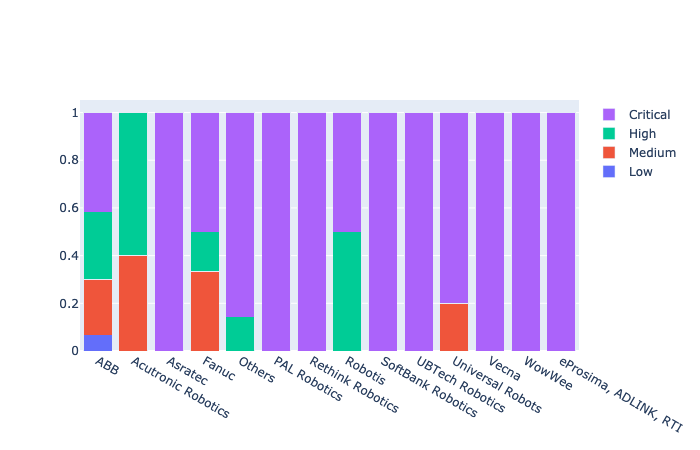}
        \caption{Non-scored flaws receive highest severity scoring.}
        \label{severity_vendor}
    \end{subfigure}
    \begin{subfigure}{.48\textwidth}
    \centering
        \includegraphics[width=\textwidth]{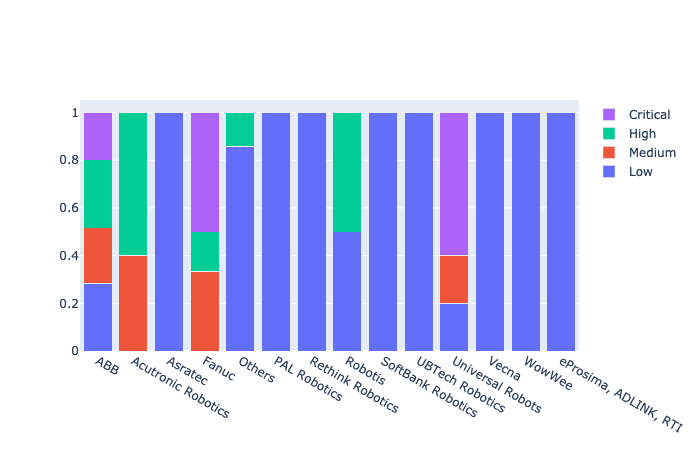}
        \caption{Non-scored flaws receive lowest severity scoring.}
        \label{severity_vendor_reversed}
    \end{subfigure}
    \centering
    \label{severity_vendor_fig}
    \caption{Entries of RVD summarizing relative proportion of Critical, High, Medium and Low scoring vulnerabilities for particular vendors in the market, according to scoring provided by CVSSv3. \ref{severity_vendor}: non-scored flaws receive the highest severity scoring. \ref{severity_vendor_reversed}: non-score flaws receive the minimum severity scoring}
\end{figure*}

In order to further investigate vulnerability severity by vendor, Figure \ref{severity_vendor} displays a barplot with the total number of flaws per vendor, relative to their severity using CVSSv3. In general, we see that except larger vendors of robotic technology, most display a vast majority of critical vulnerabilities according to CVSS. The authors argue that the reasons behind this are two-fold: first and as is common in the security domain, vulnerabilities with unclear or unreported severity are flagged with the maximum. This affects directly the flaws recorded for smaller robotic vendors where researchers simply didn't have the motivation (possibly financial) to further pursue a security assessment.  Second, the most established (and larger) vendors of robotic technology display a lower proportion of highly critical flaws. In the authors' view, while the data available is insufficient to make strong claims, a tendency to propagate the criticality percentage down can be appreciated as companies invest in security. Particularly, the case of Acutronic Robotics, which recently performed a security assessment (disclosing partially) is of relevance to draw this conclusion. This is further illustrated in Figure \ref{severity_vendor_reversed} where the severity of non-scored vulnerabilities has been reversed.\\
\newline
In any case the authors acknowledge that the information available is incomplete in all cases. It's highly likely that vendors do not disclose information about low criticality flaws which could significantly change the plot.


\section{Comparison and bias}
\label{bias}
To avoid biased conclusions, the authors assess the results with a framework by which vulnerability statistics can be judged and improved. Particularly, the method proposed by Christey and Martin \cite{christey2013buying}, maintainers of two well-known vulnerability information repositories. According to them, most of the vulnerability related statistical analyses demonstrate a serious fault in methodology and represent in some cases pure speculation aimed at justifying security budgets and spending.\\
\newline
To self-assess the results, four types of bias are considered:
\begin{itemize}
    \item \textbf{Selection bias} or what gets effectively sampled or selected for study. 
    \begin{itemize}
        \item for researchers, what software and methodology did they use to test vulnerabilities.
        \item for the database, how the database discovers and handles vulnerability disclosures from researchers to vendors.
    \end{itemize}
    
    \item \textbf{Publication bias}, what portion of the research gets published in the tickets, and perhaps most importantly, what does not get published.
    
    \item \textbf{Abstraction bias}, assigned to reflect bias in the process the database uses to assign identifiers to vulnerabilities.
    
    \item \textbf{Measurement bias}, analyse potential errors in how a vulnerability is analyzed, verified and catalogued.
    \begin{itemize}
        \item for researchers, failure to verify that a potential issue is an actual vulnerability, or in over-estimating the severity of the issue compared to how customers might prioritize the issue.
        \item for vendors, prioritization of issues to be fixed or under-estimation of the severity.
        \item for the database, how database's tickets are filled by analysts in terms of accuracy and completeness (e.g. not filling the severity or the product description details).
    \end{itemize}    
\end{itemize}

\subsection{Selection bias self-assessment}

In the current, initial phase of introducing RVD, the total number of listed vulnerabilities is still small. This also means that the conclusions here have an implicit selection bias. As the database grows and the number and scanning intensity of robots listed increases, it has to be expected that the presented picture shifts. However, it seems to be reasonable that established companies like ABB that have a broader product range than young robotics-only companies, also employ more mature engineering processes in combination with security researchers and thus have a lower number of critical vulnerabilities (percentage-wise).

\subsection{Publication bias self-assessment}
Only high impact CVEs tend to be published by researchers. Relatively few low rating vulnerabilities have been identified to be public and added to  RVD. This might be motivated by researchers in robot security publishing exclusively high impact (according to CVSS at least) findings. Also by the lack of a precise VDB (Vulnerability DataBases) to adequately triage robot vulnerabilities, or simply "finding support" these kind of systems is troublesome. The fact that robots do not solely adhere to OT, but also to IT does not facilitate the logging of vulnerabilities into VDBs.

\subsection{Abstraction bias self-assessment}
 
 RVD IDs are provided for each robotic-related flaw, regardless of the manufacturer or vendor of the robot (or robot component). For example, a given vulnerable buffer overflow in an arbitrary OpenSSL version used in robotics should be classified with a single RVD ID ticket, regardless of how many manufacturers are vulnerable to it. Both, CVEs as well as unique RVD specific iterative identifiers (RVD IDs) have been used within the schema for compatibility and de-duplication. In addition, a semi-autonomous de-duplication mechanism based in regularized logistic  regression  augmented  with human input (Active Learning) has been used to avoid duplicates of any kind. \\
 \newline
 The authors express that to the best of their knowledge, RVD has been designed to avoid being subject to abstraction bias.
 

\subsection{Measurement bias self-assessment}
This paper has found that the method for investigation is often very poorly reported in most of the robot-related vulnerabilities discovered. This paper advocates for more detailed reports, including the environment to reproduce the finding. If vulnerabilities cannot be reproduced, it is very difficult to assign an accurate severity. To this end, there is a relevant measurement bias in the existing dataset. \\
\newline
Efforts have been dedicated to reduce such bias by re-classiying, de-deduplicating and enhancing the scoring mechanism introducing a robot-specific scoring system (RVSS). RVSS scoring remains more conservative when it comes to safety aspects in robotics, as it underrates severity in data related vulnerabilities and overrates those that have or might have an effective Safety impact, in comparison to CVSS.\\
\newline
When compared to other existing robotics-related vulnerability reports (e.g. CVE-2019-13566, CVE-2019-13465 or CVE-2019-13445), entries in RVD appear more complete and provide means for its discussion, further improvement and reproduction, when possible. Still, authors acknowledge that much work is left to be done on the triage aspect and future work will focus there.


\section{Conclusions and future work}
\label{conclusion}
This paper presented the Robot Vulnerability Database along with the processes and tools for the proposed use of RVD. It has been argued that a vulnerability repository dedicated to robotics is required to account for the complexity and special characteristics of robots that are not reflected in general, IT and OT-focused  vulnerability collection projects. RVD aims to enhance existing vulnerability enumerations like CVE List with information specific to robotics. It also aims to provide information on vulnerability reproduction to increase the overall quality of the collected items. By using machine-readable formats, RVD enables a high degree of automation in processing and validation as well as querying of database entries. This greatly reduces the effort in maintenance that would otherwise become unbearable as the adoption of RVD increases.\\
\newline
A total of 110 vulnerabilities have been recorded a the time of writing. Preliminary statistics on RVD contents have been presented highlighting that already in its current initial stage, it already contains a collection of highly critical security issues across a very broad range of manufacturers. Moreover, plots show a correlation between the involvement of a manufacturer with security (funding security researchers, publishing advisories, etc.) and the spread of the severity of their related vulnerabilities.\\
\newline
As more data becomes available authors will re-iterate on their conclusions and produce more visualizations. Future work will also include, but not limited to the items listed below:

\begin{itemize}
    \item Disclose flaws of open source robot components (e.g. ROS and ROS 2)
    \item A higher degree of automation is expected to help support in the task of creating the security pipelines and maintenance of RVD
    \item Further work is foreseen to elaborate on difficulties on triage of Robot Vulnerabilities, including the reproducibility of cyber issues. 
    \item An effort is needed to foster the differentiation of OT technologies, ICS and robotics. There is still scarce data on robots.
    
\end{itemize}

The authors want to explicitly call upon the robotics and cybersecurity communities to engage with the topic of robot cybersecurity and use RVD as an essential tool to jointly proceed towards a future with secure robots.By no means the authors would like to encourage uncontrolled tampering with running robotic devices, as this may result into serious safety hazards.

\section*{Acknowledgment}

This work has partially been funded by the ROS-Industrial Quality-Assured Robot Software Components (ROSin) RedROS-I and RedROS2-I FTPs which received funding from the European Union’s Horizon 2020 research and innovation programe under the project ROSIN with the Grant Agreement No 732287. This research was also financially supported by the Spanish Government through CDTI Neotec actions (SNEO-20181238).
In particular, this action was partially supported by the regional Basque Autonomous Government's SPRI agency for the support within HAZITEK funding scheme (ZL-2019/00439) and ZABALDU internationalization actions (ZAB-00014-2019) and special thanks to the Basque Cyber Security Centre BCSC for the support in actions fostering awareness in robot Cyber Security. Last but not least, authors are grateful to the local administration Diputación Foral de Álava for the support to entrepreneurship in innovation actions (EMPREM-2019/00002). 




%

\bibliography{bibliography}  
\bibliographystyle{IEEEtran}

\appendices

\section{Terminology}
\label{appendixA}

Commonly \cite{wiki:Software_bug}, a \textbf{software bug} is an error, flaw, failure or fault in a computer program or system that causes it to produce an incorrect or unexpected result, or to behave in unintended ways.

A \textbf{software weakness} however is an error that can lead to software vulnerabilities according to the Common Weakness Enumeration \cite{weakness_cwe}. The same source fines a \textbf{software vulnerability} as a mistake in software that can be directly used by a hacker to gain access to a system or network while ISO/IEC 27001 proposes the following definition for vulnerability: bug of an asset or control that can be exploited by one or more threats.

Finally, CVE defines \cite{vulnerability_cve} a \textbf{software exposure} as a system configuration issue or a mistake in software that allows access to information or capabilities that can be used by a hacker as a stepping-stone into a system or network.

\subsection*{Discussion and interpretation}

From the definitions above, it seems reasonable to associate use interchangeably the terms bug and flaw when referring to software issues. In addition, the word weakness seems applicable to any flaw that might turn into a vulnerability however it must be noted that (from the text above) a vulnerability "must be exploited". Based on this, a clear difference can be established classifying flaws with no potential to be exploitable as bugs and flaws potentially exploitable as vulnerabilities. Orthogonal to this appear exposures which refer to misconfigurations that allows attackers to establish an attack vector in a system.\\
\newline
Beyond pure logic, an additional piece of information that comes out of researching other security databases is that most security-oriented databases do not distinguish between bugs (general bugs) and weaknesses (security bugs).\\
\newline
Based in all of the above, we interpret and make the following assumptions for RVD:
\begin{itemize}

    \item unless specified, all flaws are "security flaws" (an alternative could be a quality flaw). Flaw is used as a general term to refer to any possible security error.
    \item bug and weakness refer to the same thing and can be used interchangeably
    \item a bug is a flaw with potential to be exploited (but unconfirmed exploitability)
    \item vulnerability is a bug that is exploitable.
    \item exposure is a configuration error or mistake in software that without leading to exploitation, leaks relevant information that empowers an attacker.
\end{itemize}

\section{Relationship between RVD, CVE, and the NVD?}

To understand the relationship between these terms, we propose below some definitions:
\begin{itemize}
    \item \textbf{Robot Vulnerability Database} (RVD) is a database for robot vulnerabilities and bugs that aims to record and categorize flaws that apply to robot and robot components. RVD is created as a community-contributed and open archive of robot security flaws. It is originally created and sponsored by Alias Robotics.
    \item \textbf{Common Vulnerabilities and Exposures} (CVE) List CVE® is an archive (dictionary according to the official source) of entries—each containing an identification number, a description, and at least one public reference—for publicly known cybersecurity vulnerabilities. CVE contains vulnerabilities and exposures and is sponsored by the U.S. Department of Homeland Security (DHS) Cybersecurity and Infrastructure Security Agency (CISA). It is not a database (see official information). CVE List feeds vulnerability databases (such as the National Vulnerability Database (NVD)) with its entries and also acts as an aggregator of vulnerabilities and exposures reported at NVD.
    \item \textbf{U.S. National Vulnerability Database} (NVD) is the U.S. government repository of standards based vulnerability management data. It presents an archive with vulnerabilities, each with their corresponding CVE identifiers. NVD gets fed by the CVE List and then builds upon the information included in CVE Entries to provide enhanced information for each entry such as fix information, severity scores, and impact ratings.
\end{itemize}

\noindent RVD does not aim to replace CVE but to complement it for the domain of robotics. RVD aims to become CVE-compatible by tackling aspects such scope and impact of the flaws (through a proper severity scoring mechanism for robots), information for facilitating mitigation, detailed technical information and more.\\
\newline
When compared to other vulnerability databases, RVD aims to differentiate itself by focusing on the following:
\begin{itemize}
    \item \textbf{robotics specific}: RVD aims to focus and capture robot-specific flaws. If a flaw does not end-up applying to a robot or a robot component then it should not be recorded here.
    \item \textbf{community-oriented}: while at the time of wrriting RVD is sponsored by Alias Robotics, it aims to become community-managed and contributed.
    \item \textbf{facilitates reproducing robot flaws}: Working with robots is very time consuming. Mitigating a vulnerability or a bug requires one to first reproduce the flaw. This can be extremely time consuming. Not so much providing the fix itself but ensuring that your environment is appropriate. At RVD, each flaw entry should aim to include a field named as reproduction-image. This should correspond with the link to a Docker image that should allow anyone reproduce the flaw easily.
    \item \textbf{robot-specific severity scoring}: opposed to CVSS which has strong limitations when applied to robotics, RVD uses RVSS, a robot-specific scoring mechanism.
\end{itemize}

As part of RVD, we encourage security researchers to file CVE Entries and use official CVE identifiers for their reports and discussions at RVD.

\section{RVD schema}
\label{schema}

\begin{lstlisting}[language=Python]
SCHEMA = {
    'id': {
        'required': True,
        'oneof': [{'type': 'string'}, {'type': 'number'}],
        # 'type': 'number',
        'empty': False,
        'min': 0,
        # 'max': 100
        'default_setter':
            lambda doc: 0,
        # 'default': 0
    },
    'title': {
        'required': True,
        'type': 'string',
        'maxlength': 100,  # extend beyond 65 to cope with a few tickets
    },
    'type': {
        'required': True,
        'type': 'string',
        'allowed': ['bug', 'weakness', 'vulnerability', 'exposure'],
        'default_setter':
            lambda doc: 'bug'
    },
    'description': {
        'required': True,
        'type': 'string',
        # 'empty': False,
        # 'default_setter':
        #     lambda doc: None,
    },
    'cwe': {
        'required': True,
        'type': 'string',
        # 'oneof': [{'type': 'string'}, {'type': 'number'}],
        # # Changed in version 0.7: nullable is valid on
        # #  fields lacking type definition.
        # 'nullable': True,
        'regex': '^CWE-[0-9]*.*$|^None$',
        'default_setter':
            lambda doc: 'None'
    },
    'cve': {
        'required': True,
        'type': 'string',
        'regex': '^CVE-[0-9]*-[0-9]*$|^None$',  # CVE-2019-13585
        'default_setter':
            lambda doc: 'None'
    },
    'keywords': {
        'required': True,
        'oneof': [{'type': 'string'}, {'type': 'list'}],
        'default_setter':
            lambda doc: ''
    },
    'system': {
        'required': True,
        'type': 'string',
        'default_setter':
            lambda doc: ''
    },
    'vendor': {
        'required': True,
        'type': 'string',
        'nullable': True,
        'default_setter':
            lambda doc: None
    },
    'severity': {
        'required': True,
        'schema': {
            'rvss-score': {
                'oneof': [{'type': 'string'}, {'type': 'number'}],
                'regex': '^None$',
                'min': 0,
                'max': 10,
                'required': True,
            },
            'rvss-vector': {
                'type': 'string',
                'required': True,
            },
            'severity-description': {
                'type': 'string',
                'required': True,
            },
            'cvss-score': {
                'oneof': [{'type': 'string'}, {'type': 'number'}],
                'regex': '^None$',
                'min': 0,
                'max': 10,
                'required': False,
            },
            'cvss-vector': {
                'type': 'string',
                'required': False,
            },
        }
    },
    'links': {
        'required': False,
        'oneof': [{'type': 'string'}, {'type': 'list'}],
        # 'regex': '^None$',
        'default_setter':
            lambda doc: 'None',
    },
    'bug': {
        'rename': 'flaw'
    },
    'flaw': {
        'required': True,
        'schema': {
            'phase': {
                'required': True,
                'type': 'string',
                'allowed': ['programming-time', 'build-time', 'compile-time',
                            'deployment-time', 'runtime', 'runtime-initialization',
                            'runtime-operation', 'testing', 'unknown'],
                'default_setter':
                    lambda doc: 'unknown'
            },
            'specificity': {
                'required': True,
                'type': 'string',
                # 'allowed': ['general issue', 'robotics specific',
                #             'ROS-specific', 'subject-specific', 'N/A'],
                'default_setter':
                    lambda doc: 'N/A',
            },
            'architectural-location': {
                'required': True,
                'type': 'string',
                'allowed': ['application-specific code', 'application-specific',
                            'platform-code', 'platform code', 'ROS-specific',
                            'third-party', 'N/A'],
                'default_setter':
                    lambda doc: 'N/A',
            },
            'application': {
                'type': 'string',
                'required': True,
                'default_setter':
                    lambda doc: 'N/A',
            },
            'subsystem': {
                'type': 'string',
                'required': True,
                'regex':
                    '^(sensing|actuation|communication|cognition|UI|power).*$|^N/A$|.*',
                    # TODO: modify this value and enforce the subsystem's policies
                    # '^(sensing|actuation|communication|cognition|UI|power).*$|^N/A$',
                'default_setter':
                    lambda doc: 'N/A',
            },
            'package': {
                'oneof': [{'type': 'string'}, {'type': 'list'}],
                # 'type': 'string',
                'default_setter':
                    lambda doc: 'N/A',
            },
            'languages': {
                'required': True,
                'oneof': [{'type': 'string'}, {'type': 'list'}],
                # 'type': 'string',
                'allowed': ['Python', 'python', 'cmake', 'CMake', 'C', 'C++',
                            'package.xml', 'launch XML', 'URScript', 'shell',
                            'msg', 'srv', 'xacro', 'urdf', 'None', 'rosparam YAML',
                            'XML', 'ASCII STL', 'N/A', 'YAML', 'Package XML'],
                'default_setter':
                    lambda doc: 'None'
            },
            'date-detected': {
                ## TODO: review this and force date check
                # 'type': 'date',
                'type': 'string',
                'required': True,
                # 'coerce': 'datecoercer',
                'default_setter':
                    lambda doc: ''
            },
            'detected-by': {
                'type': 'string',
                'required': True,
                'default_setter':
                    lambda doc: ''
            },
            'detected-by-method': {
                'type': 'string',
                'required': True,
                'allowed': ['build system', 'compiler',
                            'assertions', 'runtime detection', 'runtime crash'
                            'testing violation', 'testing static',
                            'testing dynamic', 'N/A'],
                'default_setter':
                    lambda doc: 'N/A'
            },
            'date-reported': {
                'type': 'string',
                'required': True,
                'default_setter':
                    lambda doc: ''
            },
            'reported-by': {
                'type': 'string',
                'required': True,
                'default_setter':
                    lambda doc: ''
            },
            'reported-by-relationship': {
                'type': 'string',
                'required': True,
                'allowed': ['guest user', 'contributor',
                            'member developer', 'automatic',
                            'security researcher', 'N/A'],
                'default_setter':
                    lambda doc: 'N/A'
            },
            'issue': {
                'type': 'string',
                'default_setter':
                    lambda doc: '',
            },
            'reproducibility': {
                'type': 'string',
                'required': True,
                'default_setter':
                    lambda doc: '',
            },
            'trace': {
                'type': 'string',
                'required': True,
                'default_setter':
                    lambda doc: '',
            },
            'reproduction': {
                'type': 'string',
                'required': True,
                'default_setter':
                    lambda doc: ''
            },
            'reproduction-image': {
                'type': 'string',
                'required': True,
                'default_setter':
                    lambda doc: ''
            },
        }
    },
    'exploitation': {
        'required': True,
        'default_setter':
            lambda doc: '',
        'schema': {
            'description': {
                'required': True,
                'type': 'string',
                'default_setter':
                    lambda doc: ''
            },
            'exploitation-image': {
                'required': True,
                'type': 'string',
                'default_setter':
                    lambda doc: ''
            },
            'exploitation-vector': {
                'required': True,
                'type': 'string',
                'default_setter':
                    lambda doc: ''
            },
        }
    },
    'fix': {
        'rename': 'mitigation'
    },
    'mitigation': {
        'required': True,
        'schema': {
            'description': {
                'required': True,
                'type': 'string',
                'default_setter':
                    lambda doc: ''
            },
            'pull-request': {
                'oneof': [{'type': 'string'}, {'type': 'number'}],
                # 'type': 'string',
                'default_setter':
                    lambda doc: ''
            },
        }
    },
}

\end{lstlisting}

\end{document}